\documentclass[a4paper,fleqn,usenatbib]{mnras}
\usepackage{newtxtext,newtxmath}
\usepackage[T1]{fontenc}
\usepackage{ae,aecompl}
\usepackage{graphicx}	
\usepackage{amsmath}	
\usepackage{amssymb}	
\usepackage{tablefootnote}
\usepackage{supertabular}
\makeatletter
\newcommand\footnoteref[1]{\protected@xdef\@thefnmark{\ref{#1}}\@footnotemark}
\newcommand\tablefootnoteref[1]{\protected@xdef\@thefnmark{\ref{#1}}\@tablefootnotemark}
\makeatother
\usepackage{array}
\usepackage{multirow}
\usepackage[above]{placeins}
\usepackage{float}

\title[CGM of DLA Galaxies]{Characterising the Circum-Galactic Medium of Damped Lyman-$\alpha$ Absorbing Galaxies}

\author[R. Augustin et al.]{Ramona Augustin$^{1,2}$\thanks{E-mail: raugusti@eso.org},
C{\'e}line P{\'e}roux$^{2}$,
Palle M{\o}ller$^{1}$,
Varsha Kulkarni$^{3}$,
\newauthor
Hadi Rahmani$^{4}$,
Bruno Milliard$^{2}$,
Matthew Pieri$^{2}$,
Donald G. York$^{5}$,
\newauthor
Giovanni Vladilo$^{6}$,
Monique Aller$^{7}$,
Martin Zwaan$^{1}$
\\
$^{1}$European Southern Observatory, Karl-Schwarzschildstrasse 2, D-85748 Garching bei M{\"u}nchen, Germany\\
$^{2}$Aix Marseille Universit{\'e}, CNRS, LAM (Laboratoire d'Astrophysique de Marseille) UMR 7326, 13388, Marseille, France\\
$^{3}$Dept. of Physics and Astronomy, Univ. of South Carolina, Columbia, SC 29208, USA\\
$^{4}$GEPI, Observatoire de Paris, PSL Research University, CNRS, Place Jules Janssen, 92190 Meudon, France\\
$^{5}$Dept. of Astronomy and Astrophysics, Univ. of Chicago, 5640 S. Ellis Ave, Chicago, IL 60637, USA\\
$^{6}$Osservatorio Astronomico di Trieste - INAF, Via Tiepolo 11 34143 Trieste, Italy\\
$^{7}$Georgia Southern University, Dept. of Physics, Statesboro, GA 30460, USA\\
}

\date{Accepted XXX. Received YYY; in original form ZZZ}

\pubyear{2018}

\begin{document}
\label{firstpage}
\pagerange{\pageref{firstpage}--\pageref{lastpage}}
\maketitle

\begin{abstract}
Gas flows in and out of galaxies through their circumgalactic medium (CGM) are poorly {constrained} and direct observations of this faint,  {diffuse} medium remain challenging.
We use a sample of five $z$ $\sim$ 1-2 galaxy counterparts to Damped Lyman-$\alpha$ Absorbers (DLAs) to combine data on cold gas, metals and stellar content of the same galaxies. 
We present new HST/WFC3 imaging of these fields in 3-5 broadband filters and characterise the stellar properties of the host galaxies. 
By fitting the spectral energy distribution, we measure their stellar masses to be in the range of log($M_*$/$\text{M}_{\odot}$) $\sim$ 9.1$-$10.7.
Combining these with IFU observations, we find a large spread of baryon fractions inside the host galaxies, between 7 and 100 percent. 
Similarly, we find gas fractions between 3 and 56 percent.
Given their star formation rates, these objects lie on the expected main sequence of galaxies. 
Emission line metallicities indicate they are consistent with the mass-metallicity relation for DLAs.
We also report an apparent anti-correlation between the stellar masses and $N$(\ion{H}{i}), which could be due to a dust bias effect or lower column density systems tracing more massive galaxies. 
We present new ALMA observations of one of the targets leading to a molecular gas mass of log($M_{\rm mol}$/M$_{\odot}$) < 9.89.
We also investigate the morphology of the DLA counterparts and find that most of the galaxies show a clumpy structure and suggest ongoing tidal interaction. 
Thanks to our high spatial resolution HST data, we gain new insights in the structural complexity of the CGM.

\end{abstract}




\section{Introduction}

Tremendous progress has been made over the last decade in establishing a broad cosmological framework in which galaxies and large-scale structure develop hierarchically over time, because of gravitational instability of material dominated by dark matter (e.g., \citealt{springel03}). 
The next challenge is to understand the physical processes of the formation of galaxies and their interactions with the medium surrounding them. 
However, details of the cycling of gas into, through, and out of galaxies are currently poorly {constrained}. 
The missing link is information about gas in the circum-galactic medium (CGM).
The CGM is loosely defined in the literature {(e.g. \citealt{tumlinson17})} but we use it as a description of the vicinity around galaxies where all the gas flows and  interactions between the galaxy itself and the primordial gas from the cosmic web take place.
The extent of the CGM can reach out to $\sim$ 300 kpc from the center of the galaxy \citep{steidel10,prochaska11}.
The CGM is believed to be both the repository of the inflowing gas and the receptacle of the feedback of energy and metals generated inside the galaxy \citep{tumlinson17}.
Since outflows can also leave the halo and mix with the primordial gas in the cosmic web, we further restrict the definition of what we consider CGM to the halo in which the gas is gravitationally bound to the central galaxy.
{Results from the FIRE (Feedback In Realistic Environments) simulation \citep{muratov15,muratov17,angles-alcazar17} suggest that the CGM at late times is mainly a repository for the metal enriched gas from outflows, and in this stage forms the major contribution to feeding the galaxy.
The observational signature of this transition from 'infall dominated' to 'closed box evolution', i.e. a steepening of the cosmic evolution of the mass to metal relation, has already been reported to take place at the time of Cosmic Noon \citep{moller13}.} 
\citet{rahmati14} {used cosmological simulations at z=3 to show} that at impact parameters larger than the virial radius of a single galaxy, galaxy clustering starts to play a role and that single absorption features are associated to groups of galaxies.
{They also find that depending on the star formation rate of the central galaxy, clustering can already become significant even within the virial radius.}

To reach understanding of the interplay of gas flows and galaxy evolution, it is essential to relate {the cold gas and metals in the CGM to the stellar content of the central galaxy.}
The cold gas and metals in galaxies can be efficiently probed using absorption lines in the spectra of background quasars. 
{These quasar absorbers offer powerful redshift-independent tools to investigate galaxy evolution.}
The quasar absorbers with the strongest absorption lines are the Damped Lyman-$\alpha$ systems (DLAs).
They are defined to have neutral gas column density log($N$(\ion{H}{i})/cm$^{-2}$)$>$20.3, matching the detection threshold of \ion{H}{i} in local disks \citep{wolfe86}. 
\cite{peroux03} showed that by extending this definition to log($N$(\ion{H}{i}))$>$19.0, all the significant contributors to the neutral gas mass density would be taken into account (see also \citealp{zafar13}). 
They coined the term sub-Damped Lyman-$\alpha$ systems (sub-DLAs) for these absorbers {with lower column densities}. 
{These DLAs and sub-DLAs are, due to their strong absorption features, probably linked to galaxies and are the preferred targets to study the CGM in absorption.}

{
While the number of quasar absorption line systems now amount to hundreds of thousands (e.g. \citealt{prochaska05,york06,noterdaeme09,noterdaeme12,zhu13,khare14,pieri14,quiret16}), the number of counterpart, visible objects attributed to DLAs (and thus termed "DLA galaxies") remains small \citep{krogager17}. 
This implies that many counterparts remain under the detection limit (see e.g. \citealt{peroux11a}) and makes the analysis of those counterparts that are being discovered even more precious.
}

{
Traditionally, people have used imaging and follow-up spectroscopy \citep{steidel97} or narrow-band imaging \citep{rauch08,ogura17}.
}
{To find the counterparts of DLAs in a more systematic way, a strategy using triangulation of the X-Shooter slit has been developed. 
This observational strategy resulted in many new discoveries of DLA galaxies \citep{noterdaeme12a,krogager17,zafar17}.
}
The use of IFUs such as SINFONI and MUSE as well as KCWI has turned out to be very powerful in detecting faint galaxies in the vicinity of bright quasars \citep{bouche12,peroux11a,rahmani18,fumagalli17,bielby17,rubin17}. 
This technique allows one to identify emission lines which in turn give constraints on the redshift, metallicity and star formation rates of the host galaxies. 
However, it provides only limited information on the stellar continuum of the host galaxies{, depending on the wavelength coverage of the IFU and the detection limit of the instrument}.
{Recent studies show that absorbers are generally not to be associated to isolated galaxies. Some absorbers are found towards groups of galaxies, suggesting that they probe some common inter-group material \citep{peroux16,rahmani18,klitsch18}.}

{With the counterpart galaxies that were detected, there} has been much progress constraining their physical properties.
{The best way to constrain these physical properties is to infer and analyse scaling relations between the gas in absorption and the emission properties from the host galaxy.
These scaling relations include a luminosity-metallicity relation \citep{moller04} or the mass-metallicity relation \citep{ledoux06,prochaska08,arabsalmani18}.}
Especially the mass-metallicity relation of DLA counterparts, and its redshift evolution, has been studied in great detail recently (e.g. \citealp{moller13,neeleman13,christensen14}).
{In these relations, the metallicity is inferred from the absorption features in the quasar spectrum and found to be in correlation with the luminosity or the mass of the galaxy counterpart.
}
However, due to the proximity of the usually faint DLA host galaxy to a bright background quasar, direct imaging and discovery of the DLA hosts has proven to be challenging (e.g. \citealp{lebrun97,moller02,chen03,rao03,kulkarni06,fumagalli14,fumagalli15}).
According to the known scaling relations, metal rich, and therefore luminous and massive systems, have the highest chance of being detected as galaxy counterparts (e.g. \citealp{fynbo10,peroux11a,bouche12,krogager13,zafar17,rudie17}).
{Even though there has been enormous progress in developing these detection techniques, the total number of detected (sub-)DLA counterparts remains relatively small and further searches for counterparts are needed in order to fully understand quasar absorbers.}

To constrain the stellar content of DLA counterparts, we acquired new broad-band HST/WFC3 observations with high spatial resolution to image five $z\sim$1-2 DLAs in three to five different filters. 
These new data provide the possibility to study the stellar continuum light of Lyman-$\alpha$ absorbing galaxies. 
{We combine our data with archival data and estimate the galaxies' masses}, test their scaling relations and use the high resolution of the data to investigate the morphology of the counterparts.

We structure the presentation of our work as follows: 
in section 2 we present the HST data set of our sample of galaxies and in section 3 the new ALMA observations.
We show the details of our data analysis in section 4.
In section 5 we present and discuss our results, and we describe our conclusions in section 6.

\begin{figure*}
	\includegraphics[width=1.5\columnwidth]{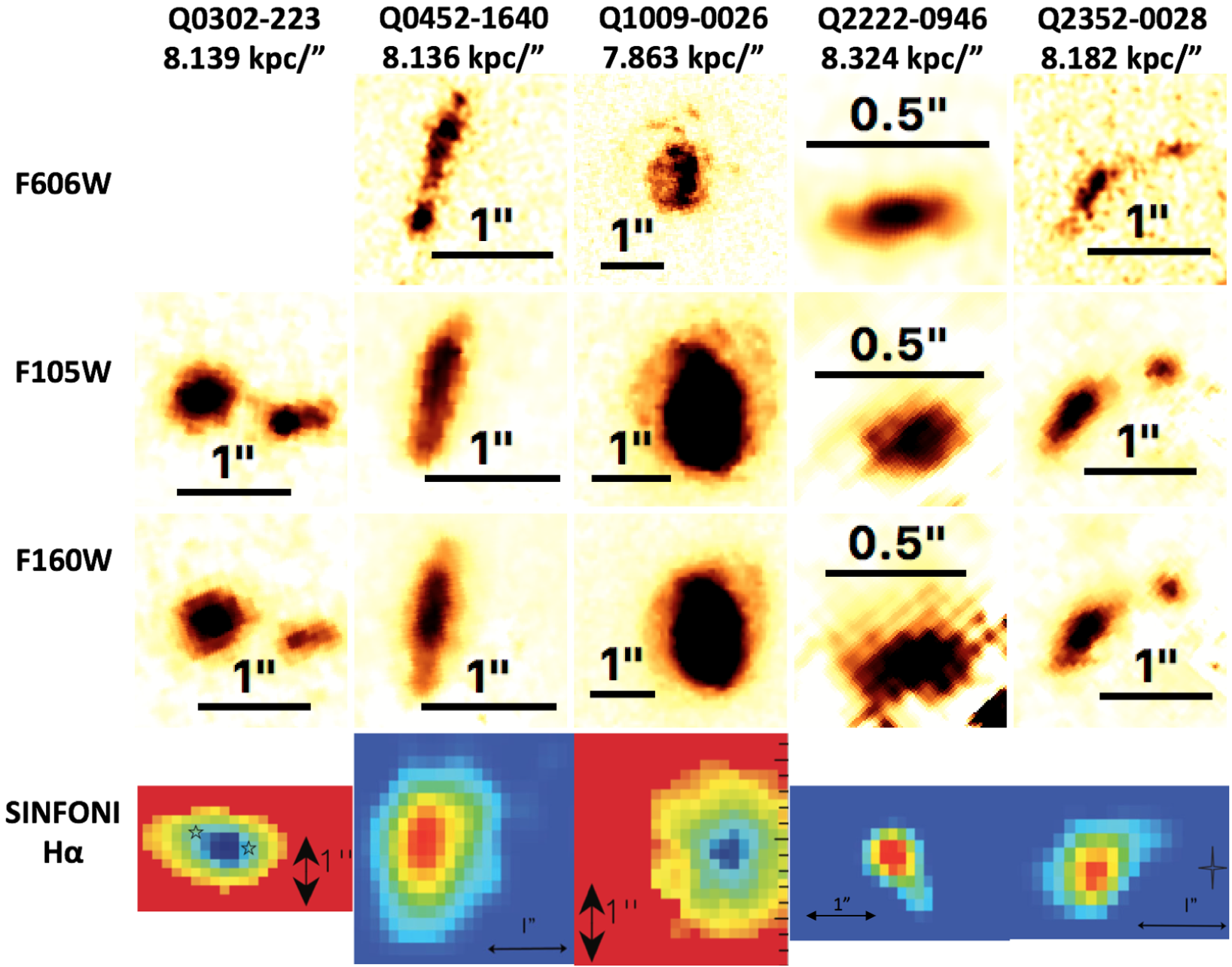}
    \caption{HST/WFC3 visible and infrared broad-band images of the stellar continuum of the galaxies of our sample after PSF subtraction. We present the objects that we find at the position where we had previously found H-$\alpha$ emission at the redshift of the absorber.
    North is up and east is left. 
    {A scale bar is shown in each panel and the corresponding physical scales are given on top of each column.} The field of Q0302$-$223 was not observed with WFC3/F606W because archival WFPC2/F450W and F702W data were already available. All the Lyman-$\alpha$ absorbing galaxies are clearly detected in all bands, even before quasar PSF subtraction.
    Note that the images in F105W and F160W for Q2222-0946 look blurred due to the proximity (0.7") of this object to the quasar sight line {because it is affected by the subtraction residuals}.
    {In the last row we also show the SINFONI H$\rm \alpha$ maps of the corresponding objects. These figures are adapted from \citet{peroux11b} and \citet{peroux12}.}
    }
    \label{fig:sample}
\end{figure*}

\section{HST Broad-band Observations of Lyman-$\alpha$ Absorbing Galaxies}

\subsection{A Sample of five DLA Counterparts}

 The data set is composed of five $z\sim$1-2 DLAs with well-determined gas and metal properties and galaxy counterparts identified from ground-based observations (see table \ref{tab:knowndata}). 
{The properties of these galaxies were previously published in a series of papers \citep{peroux11a,peroux11b,peroux12,peroux13}.
\citet{peroux11a} presented the VLT/SINFONI IFU survey in which these sources were included.
They targeted quasar fields with known DLAs in order to look for H-$\rm \alpha$ emission in these fields at the redshift of the absorption, and thereby identified galaxy counterparts.
In \citet{peroux11b} the kinematics of two of these counterparts were analysed.
\citet{peroux12} introduced three more counterpart detections and summarised the emission metallicites from the N2-parameter, which is based on the \ion{N}{II} $\rm \lambda$6585 / H-$\rm \alpha$ ratio \citep{pettini04}.
 Later, \citet{peroux13} provided dynamical masses, gas masses and halo masses from 3D morphokinematics modelling for these DLA galaxies and \citet{peroux14} used VLT/X-Shooter data to estimate the metallicity of the \ion{H}{II} gas also from R23 \citep{kobulnicky04}.
Their parameters that are relevant for this work are presented in table \ref{tab:knowndata}, and include accurate sky positions, impact parameters, the redshift of the DLA galaxy, emission line metallicities, kinematics and H-$\rm \alpha$ star formation rates.
}

{
The SINFONI data of these objects did not reveal the galaxies in continuum, only in emission lines and in order to analyse the stellar continuum, an HST follow-up has been undertaken.
}

\begin{table*}
	\centering
	\caption{Physical properties of absorbing galaxies observed with SINFONI \citep{peroux11a, peroux11b}. {In the first column we describe the (sub-)DLA counterpart by the quasar field and the clump ID as defined later in figure \ref{fig:label}.} "$\delta$" refers to the angular separation between the quasar and the galaxy, "$b$" refers to the impact parameter in kpc. {$z_{\text{DLA}}$ gives the redshift of the absorber and log($N$(\ion{H}{i})) its \ion{H}{I} column density. The SFR was determined from H-$\rm \alpha$.} We note that no extinction corrections have been applied to the star formation rate (SFR) estimates. The absorption metallicity corresponds to the neutral gas abundance derived from Zinc, assumed to be undepleted on to dust, while the metallicity in emission is from the \ion{H}{II} regions within the galaxies {using N2 (upper row) or R23 (lower row)}. {We also show the derived maximum velocities and mass components.}}
	\label{tab:knowndata}
	\begin{minipage}{200mm}
	\resizebox{.89\textwidth}{!}{
	\begin{tabular}{ccccccccccc}
		\hline
		Quasar Field &$\delta$ ["]   & $z_{\text{DLA}}$ & log($N$(\ion{H}{i})) & SFR (H-$\rm \alpha$) & log($X/H$) & {log($X/H$) [N2]} & $V_{\rm max}$  & log($M_{\rm dyn}$) & log($M_{\rm halo}$)  &log($M_{\rm gas}$)\\
		(Object ID) & $b$ [kpc] & &[cm$^{-2}$] & [$\text{M}_{\odot}$/yr] & (absorption) & {log($X/H$) [R23]}  & [km/s]  & [$\text{M}_{\odot}$] & [$\text{M}_{\odot}$] & [$\text{M}_{\odot}$] \\
		\hline
		Q0302$-$223\footnote{\label{peroux11btab} \citep{peroux11b}} & 3.16   &1.009  & 20.36$\pm$0.04\footnote{\label{rao06tab} \citep{rao06}} & 1.8$\pm$0.6 & $-0.51 \pm 0.12$ & < $-0.06$  & 11 & 10.3 & - &9.1\\
		(3+4) &25&&&&& $0.07 \pm 0.23$&&&&\\
		Q0452$-$1640\footnote{\label{peroux11atab} \citep{peroux11a}} & 2.00   &1.007   & $20.98^{+0.06}_{-0.07}$\footnoteref{rao06tab} & 3.5$\pm$1.0 & $-0.96 \pm 0.08$ & $ -0.26 \pm 0.1$  & 100 & 10.6 & 12.8 &9.2\\
				(1+2) & 16&&&&& $-0.7 \pm 0.04$&&&&\\
		Q1009$-$0026\footnoteref{peroux11btab} & 5.01  & 0.887  & $19.48^{+0.05}_{-0.06}$\footnoteref{rao06tab} & 2.9$\pm$1.0 & $+0.25 \pm 0.06$ & $+0.04 \pm 0.80 $  & 250 & 10.9 & 12.6 &9.2\\
				(1) &39&&&&& $--$&&&&\\
		Q2222$-$0946\footnoteref{peroux11atab} & 0.70 & 2.354  & 20.50$\pm$0.15\footnote{\label{prochaska07tab}\citep{prochaska07}} & 17.1$\pm$5.1 & $-0.46 \pm 0.07$ & $-0.3 \pm 0.19$\footnote{\label{krogager13tab1}\citep{krogager13}}   & 20  & 9.8 & - &9.7\\
						(1)  &6&&&&& $-0.25 \pm 0.19$\footnoteref{krogager13tab1}&&&&\\
		Q2352$-$0028\footnoteref{peroux11atab} & 1.50  &1.032  & $19.81^{+0.14}_{-0.11}$\footnoteref{rao06tab} & 1.3$\pm$0.6 & $ < -0.51 $ & $ -0.26 \pm 0.3$  & 140  & 10.4 & 11.8 &8.8\\
						(1+2) &12&&&&& $--$&&&&\\
		\hline
	\end{tabular}
	}
	\end{minipage}
\end{table*}

\begin{table*}
	\centering
	\caption{New and archival HST observations of the five quasar fields. The first column gives the ID of the quasar field (consistent with the IDs in previous literature, e.g. \citealt{peroux11a}); the second, the quasar redshift; the third and fourth columns, the coordinates (J2000) of the quasar; the fifth and seventh columns, the filters of the broad-band observations; and the sixth and eighth columns give the number of exposures and their respective exposure times in seconds. }
	\label{tab:dataset}
	\begin{minipage}{180mm}
	\centering
	\begin{tabular}{cccccccccc} 
		\hline
		Quasar Field &$z_{\rm QSO}$& R.A. & Dec & UVIS & $n \ \times \ \Delta t \  [{\rm s}]$ & IR & $n \ \times \ \Delta t \  [{\rm s}]$  \\
		\hline
		Q0302$-$223 &1.409 &03:04:49.79 & $-$22:11:53.17 & WFPC2/F450W \footnote{PI: Bergeron, Proposal ID: 5351, \citet{lebrun97} \label{5351}} & 4 $\times$ 500 & WFC3/F105W \footnote{PI: P\'eroux, Proposal ID: 13733, this work \label{13733}} & 4 $\times$ 653\\
		 &&  &  & WFPC2/F702W \footnoteref{5351} & 6 $\times$ 600 & WFC3/F160W \footnoteref{13733}& 4 $\times$ 653\\
		Q0452$-$1640 &2.679 & 04:52:14.19 & $-$16:40:17.07 & WFPC2/F814W \footnote{PI: Surdej, Proposal ID: 5958, \citet{surdej97} \label{5958}}& 4 $\times$ 400 & WFC3/F105W \footnoteref{13733} & 4 $\times$ 653\\
		& &  &  & WFC3/F606W \footnoteref{13733} & 4 $\times$ 626 & WFC3/F140W \footnote{PI: Erb, Proposal ID: 12471 \label{12471}}& 4 $\times$ 203\\
		& &  &   &  &  & WFC3/F160W \footnote{PI: Law, Proposal ID: 11694, \citet{erb10} \label{11694}}& 9 $\times$ 899 \\
		Q1009$-$0026 &1.242& 10:09:30.46 & $-$00:26:18.97 & WFC3/F606W \footnoteref{13733}& 4 $\times$ 624 & WFC3/F105W \footnoteref{13733}& 4 $\times$ 653\\
		& &  &   &  &  & WFC3/F160W \footnoteref{13733} & 4 $\times$ 653\\
		Q2222$-$0946 &2.927 & 22:22:56.11 & $-$09:46:36.35 & WFC3/F606W \footnote{PI: Fynbo, Proposal ID: 12553, \citet{krogager13} \label{12553}} & 4 $\times$ 629 & WFC3/F105W \footnoteref{12553} & 4 $\times$ 653\\
		& &  & &  &  & WFC3/F160W \footnoteref{12553} & 4 $\times$ 653  \\
		Q2352$-$0028 &1.624& 23:52:53.54 & $-$00:28:50.57 & WFC3/F606W \footnoteref{13733}& 4 $\times$ 624 & WFC3/F105W \footnoteref{13733}& 4 $\times$ 653\\
		& &  &  &  &  & WFC3/F160W \footnoteref{13733}& 4 $\times$ 653\\
		\hline
	\end{tabular}
	\end{minipage}

\end{table*}

\begin{figure}
	\includegraphics[width=1.0\columnwidth]{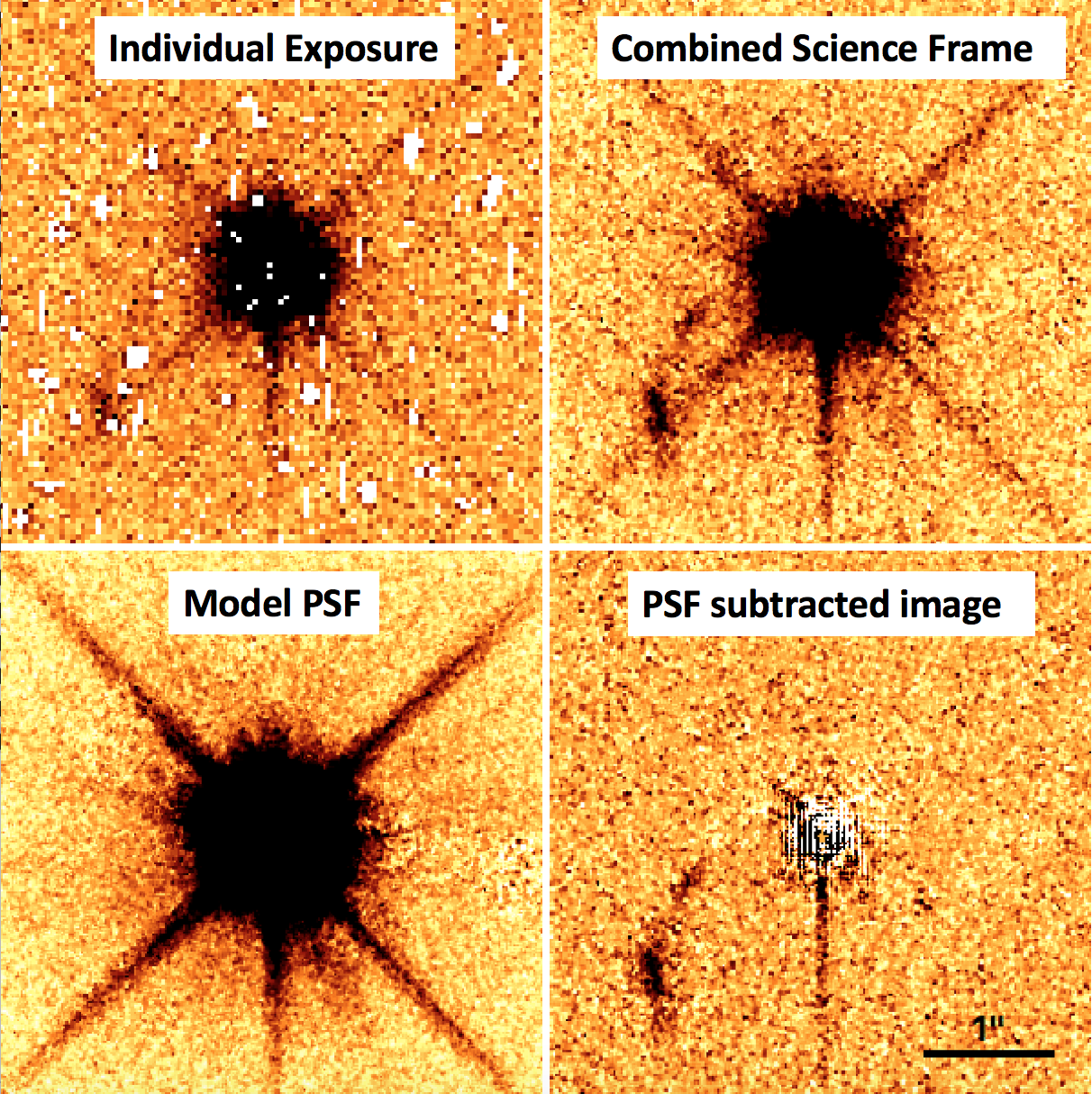}
    \caption{{A worst case example} of our quasar PSF subtraction method in the case of Q2352$-$0028 in WFC3/F606W. Top left: cutout of one of the four individual exposures. Bad pixels and cosmic rays are masked {(white spots)}. Every frame is shifted to correct for sub-pixel offsets and resample (10$\times$10 pixels) for alignment. Top right: Median stack of all the aligned individual images. Bottom left: Model PSF obtained from median stacking the other fields {(Q0452$-$1640, Q1009$-$0026 and Q2222$-$0946)} in our sample in the appropriate filter after masking objects which do not belong to the quasar PSF {and scaling the them to one another}. Bottom right: Resulting final PSF-subtracted image {of Q2352$-$0028 after scaling the model PSF in the lower left panel to the QSO in Q2352$-$0028}. This method provides in a good correction of the quasar PSF. In particular, the diffraction spikes, which are spreading over several arcsec and as such are more likely to affect the detection of the absorbing galaxies, are robustly removed.}
    \label{fig:psfsub}
\end{figure}

\subsection{HST WFPC2 and WFC3 Observations}

To study the stellar properties {such as the stellar mass} of these galaxies, {which requires knowledge of the underlying stellar continuum,} we combined imaging data from the HST archive with new WFC3/UVIS and IR channel observations. 
Thereby, we obtain broad-band fluxes and the morphology of all the objects in the sample. 
{We use the high spatial resolution of HST data to resolve the morphology of the (sub-)DLA counterparts.}

The new data were observed during Cycle 22 (ID: 13733; PI: P{\'e}roux) with the Wide Field Camera 3 in both the optical (UVIS) and infrared (IR) detectors, using the three broadband filters F606W, F105W and F160W. The observations took place between 13 January and 13 October 2015. We aimed at setting the roll-angle of the telescope such that the {known} galaxy counterpart of the quasar absorbers lies at 45 degrees from the diffraction spikes of the Point Spread Function (PSF). We use a dithering pattern in four individual exposures to help with removal of cosmic rays and hot pixels. The UVIS observation was taken using the WFC3-UVIS-DITHER-BOX pattern. The two observations with the IR detector were taken using the WFC3-IR-DITHER-BOX-MIN pattern providing an optimal 4-point sampling of the PSF. A summary of the observational set-up is given in table~\ref{tab:dataset}. A minimum of three broad-band filters is available for each field to probe either side of their 4000\AA\ Balmer break. Using additional data available in the archive, we end up with 4 filters for Q0302-223 and 5 filters for Q0452-1640. 
The available F606W, F105W and F160W images are displayed in figure~\ref{fig:sample}. 

\subsection{HST/WFC3 Data Processing}

We perform the data reduction with the {\it calwf3} pipeline.
This includes corrections for bias and dark as well as flat fielding. 
We then multiply each individual exposure by the pixel area map provided by the HST/WFC3 photometry website\footnote{http://www.stsci.edu/hst/wfc3/pam/pixel\_area\_maps} to do the flux calibration in the individual science frames.
All bad pixels, including cosmic rays and saturated pixels are masked using the data quality file that comes with each science frame (Fig. \ref{fig:psfsub}, upper left).
We create a subpixel grid on the individual exposures of 10$\times$10 per pixel and shift the individual frames on top of each other to align them.
Thereby we correct for the offset from the dithering.
The individual images are sky subtracted and then combined in a median stack after alignment.
This stacked image results in the science image that we are using for our analysis (figure \ref{fig:psfsub}, upper right).\\

\subsection{Quasar Point Spread Function Subtraction}

In order to detect the faint continuum emission from the foreground
Lyman-$\alpha$ absorbing galaxy near the bright quasar, we need to subtract the
quasar continuum. 
To this end, we create a model of the point spread function (PSF) directly from our data for each QSO field.
The QSO PSF is dependent on the filter in which the observations were taken and the position of the QSO on the detector plane.
Since we have taken our observations in the same three filters for every field and positioned the QSO close to the center of the detector plane, we can combine different quasar images in identical filters to create a model PSF.
We proceed as follows:
For each QSO field, we combine all other {QSO fields} in the identical filter in a median stack (figure \ref{fig:psfsub}, lower left).
Before the stacking we take care to mask all the objects that do not belong to the QSO PSF {(e.g. small galaxies, artifacts, etc.)} and subtract the sky.
We also scale the individual PSFs so that their flux profiles in the outer wings match.
This resulting model PSF is then scaled to the quasar field under study.
We take care to combine all fields but the one to be PSF subtracted, to reveal objects under the QSO PSF.
Using this model, we apply the PSF correction on the science frame (figure \ref{fig:psfsub}, lower right).
This method results in a good correction of the quasar PSF. In particular, the diffraction spikes which are spreading over several arcsec and as such are more likely to affect the detection of the absorbing galaxies are robustly removed. 
For the two archival images in Q0452$-$1640 (F814W and F140W) we simply use an isolated star in the field as the model PSF.

\begin{figure*}
	\centering
	\includegraphics[width=1.5\columnwidth]{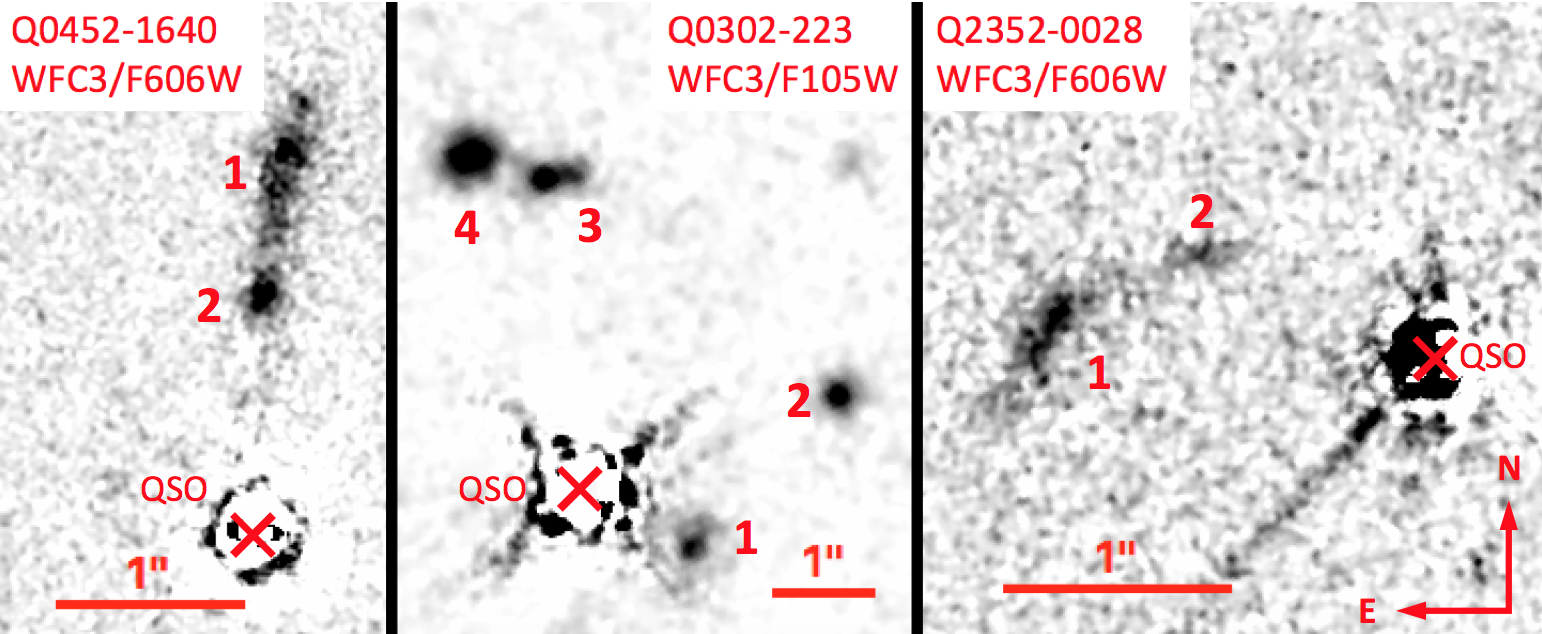}
    \caption{Nomenclature of the three fields showing several absorbing-galaxy clumps. In these fields, the enhanced spatial resolution provided by the HST/WFC3 images reveal distinct structure within the counterpart objects. North is up and East is left. The quasar PSF has been subtracted.
    We note that in the field of Q0302$-$223, previous HST/WFPC2 observations had already provided evidence of two separate objects associated with the absorbing-galaxy \citep{lebrun97,peroux11b}. 
    As we do not have spectral information on the individual clumps we add up the fluxes of the individual clumps that correspond to the DLA counterpart as seen in the ground based H-$\alpha$ detection.
    The tail in the bottom right corner of the Q2352$-$0028 field is an artefact due to the saturation of the QSO that could not be removed by the PSF subtraction.
    {We did not mask the QSO residuals which leads to some bright and dark spots around the marked QSO positions.}
}
    \label{fig:label}
\end{figure*}

\begin{table*}
	\centering
	\caption{AB magnitudes of the DLA counterpart galaxies. A description of how we obtain these magnitudes is given in section 4.1. The measurements serve as input to SED stellar continuum fitting of the Lyman-$\alpha$ absorbing galaxies.
	}
	\label{tab:magnitudes}
	\begin{minipage}{180mm}
	\centering
	\begin{tabular}{ccccc} 
		
		\hline
		Quasar Field  & UVIS & AB $mag$ & IR & AB $mag$  \\
		{(Object ID)} & & & & \\
		\hline
		Q0302-223   & F450W & 24.01 $ \pm $ 0.34\footnote{\citep{chen03}\label{chen03tab}} & F105W & 22.6 $ \pm $ 0.2  \\
		  {(3+4)} & F702W & 23.25 $ \pm $ 0.06\footnoteref{chen03tab} & F160W & 22.4 $ \pm $ 0.2 \\

		Q0452-1640  & F606W & 23.5 $ \pm $ 0.4 & F105W & 22.9 $ \pm $ 0.2  \\
		  {(1+2)} & F814W & 23.2 $\pm$ 0.3 & F140W & 23.0 $ \pm $ 0.2 \\
		  &  & & F160W & 23.2 $ \pm $ 0.3 \\
		 
		Q1009-0026   & F606W & 22.1 $\pm$ 0.2 & F105W & 20.30$ \pm $ 0.07 \\
		   {(1)} & & & F160W & 19.82 $\pm$ 0.06 \\
		 
		Q2222-0946   & F606W & 24.29 $ \pm $ 0.04\footnote{\label{krogager13tab}\citep{krogager13}}  & F105W & 24.51 $ \pm $ 0.21\footnoteref{krogager13tab} \\
		   {(1)}& & & F160W & 23.94 $ \pm $ 0.19\footnoteref{krogager13tab} \\

		Q2352-0028  & F606W & 24.5 $\pm$ 0.6 & F105W & 23.4  $ \pm $ 0.3  \\
		   {(1+2)} & & & F160W & 23.2 $\pm$ 0.3  \\
		\hline
	\end{tabular}
	\end{minipage}

\end{table*}

\section{New ALMA Observations}

The field of Q2222$-$0946 was observed with ALMA to cover the CO(3-2) emission line at the redshift of the $z_{\rm abs}$=2.354 absorber in band 3 on the 10 August 2016 (under program 2015.1.01130.S, PI: C. P\'eroux) using a compact antenna configuration. 
One of the four spectral windows was centered at the redshifted CO(3-2) line frequency of 103.22326 GHz, while the three other spectral windows were used for continuum observations of the field. 
The blazars J2224$-$1126 and J2148$+$0657 were used as calibrators. 
Full imaging pipeline products are generated by the observatory using the Common Astronomy Software Applications (CASA) software package version 4.7.0. 
The final beam size is 1.04$\times$0.78 arcseconds. 
No emission lines are detected at the expected frequency. 
The resulting RMS at the observed frequency is 0.139 mJy in 150 km/s channel width.
Assuming a line width of 200 km/s, we derive a 5-sigma limit on the integrated flux density of <140 mJy km/s. 
Converting to CO(1-0), we get a flux density of <58 mJy km/s assuming a Milky-Way like J-transition ratio \citep{carilli13}.
{In line with previous works, we choose to use a galactic H2-CO conversion factor of 4.6 M$_{\odot}$ (K km/s pc$^2$)$^{-1}$, although this assumption has been challenged by recent findings (e.g. \citealt{klitsch18}).} 
The corresponding CO luminosity is $1.7 \times 10^9$ (K km/s pc$^2$)$^{-1}$ leading to a 5-sigma limit on the molecular mass of log($M_{\rm mol}$/M$_{\odot}$) < 9.89.

\section{Analysis}

\subsection{Magnitude Measurements}

We perform aperture photometry to determine the magnitudes of the DLA counterparts.
We carefully choose elliptical apertures (major axis lengths between 0.4 and 1.7 arcsec, depending on apparent size of the galaxy) around the objects we identify as DLA counterparts.
The visible counterpart should lie completely within the aperture.
If there are several individual sub-structures (hereafter referred to as 'clumps' {and labeled in Figure \ref{fig:label}}), we pay attention to include them all {when} detected in all filters.
{What} we identify as the DLA counterpart may not be a single object but actually several clumps that we treat as a single galaxy due to their spatial proximity and lack of further spectral information.
Also, because the area in the center of the QSO is highly contaminated, {we take care} not to include such pixels in the aperture.
In some cases, when {faint and minor} clumps show in the infrared that were not detected in the optical filter, we choose to only analyse those {major} clumps that we can see in all filters.
Therefore we will either mask those objects in the infrared or choose such an aperture that those clumps are {excluded}.
We convert the flux within this aperture to an AB magnitude using the respective zeropoints of each filter. 
The error on the magnitude within this aperture is dominated by the Poisson noise on the pixels.

Since the galaxies are extended objects we need to correct these magnitudes for the missing flux of the galaxies outside the aperture.
To do this we model Sersic profiles to each of the visible clumps and galaxies, using the software GALFIT \citep{peng02}.
Assuming that the galaxy profile looks the same in all filters outside the chosen aperture, we do this modeling {in the filter with the best spatial resolution} (F606W for all but Q0302-223, where we choose F160W).
This model is then scaled to the resolution and pixel sampling of the other filters and used to determine the IR aperture correction.
Indeed, clumps that are not detected in the optical might show in the infrared.
{These could be unrelated to the galaxy we are studying and therefore have an effect on the fit, but we find that including these regions has no impact on the derived magnitudes within our conservative errors.
The clumpy residuals shown in Figure \ref{fig:morphologyandgalfit} are not related to such "infrared-only" clumps and instead show the deviation of the observed galaxies from the ideal Sersic model (see section 5.4).
We used the best fit of the model excluding the infrared-only clumps} to determine the integrated flux outside the previously chosen aperture.
This gives us the aperture correction.

The error on this aperture correction is estimated from 1000 random realisations of the model galaxy.
The Sersic profile that we fit depends on seven parameters (the center (x,y), the integrated magnitude, the effective radius, the Sersic index, the ellipticity and the position angle).
Assuming that they are independent - in reality they are not, but for our purposes this approximation is good enough - we let each parameter vary within a normal distribution around its best value and with their fitting errors as the sigma.
Taking random values within these distributions we create the 1000 realisations of the galaxy model.
Again, we measure the flux outside the chosen aperture and determine the standard deviation of this flux.
This gives us the error on the aperture correction.
Adding now the aperture correction and its error onto the flux within the aperture we obtain the magnitudes as given in table \ref{tab:magnitudes}.
 
{For comparison we} also use the GALFIT modeling on all filters to compute the magnitudes and find them to be consistent within the errors with the magnitudes that we obtain from aperture photometry. 
One object in our sample (Q2222-0946) overlaps with the study from \citet{krogager13}. 
Although we use their estimates for the {remainder} of our analysis, we perform our magnitude measurement procedure also on their original data and find that the magnitudes for this object also agree within the errors.

\subsection{Morphology Characterisation}

\begin{figure}
	\includegraphics[width=1.0\columnwidth]{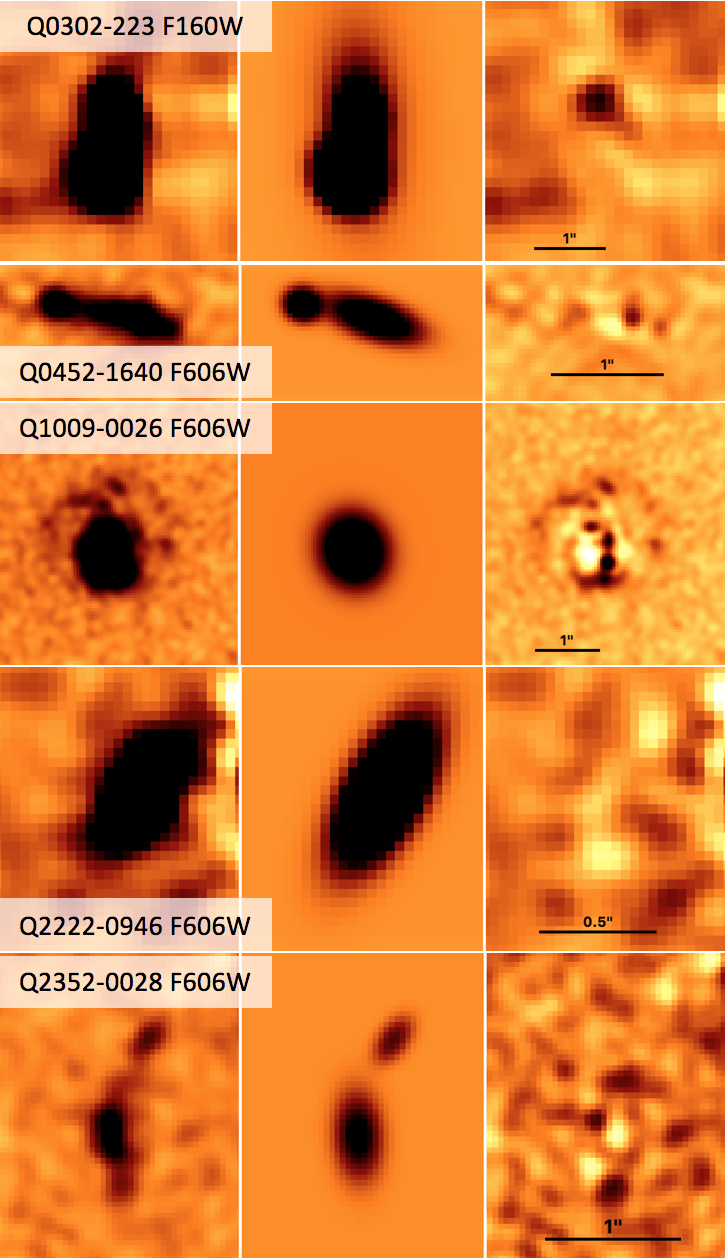}
	    \caption{Galaxy models with GALFIT. These models were used to determine the morphological parameters given in table \ref{tab:galfitparameters}. These models were also used to determine the aperture correction for the magnitudes. From left to right: the observed, modeled and residual from our GALFIT modeling. The enhanced spatial resolution provided by HST imaging reveals the complexity of the morphology of these objects. See section 5.4 for the discussion of the morphology of each of these objects.  }
    \label{fig:morphologyandgalfit}
\end{figure}

In addition, the GALFIT models {(see figure \ref{fig:morphologyandgalfit})} allow us to characterise the morphology of the DLA counterparts.
In three of the targeted fields (Q0302$-$223, Q0452$-$1640 and Q2352$-$0028), the enhanced spatial resolution provided by the HST/WFC3 images reveals distinct structure within the absorbing galaxies. 
Figure~\ref{fig:label} provides the nomenclature for each of these clumps. 
For the other two fields, no additional structure is found within the galaxy counterpart {itself, although we find quite a lot of structure around the quasar in the field of Q1009-0026 (see figure \ref{fig:1009field})}.
We note that in the field of Q0302$-$223, previous HST/WFPC2 observations had already provided evidence of two separate objects associated with the absorbing-galaxy \citep{lebrun97,peroux11b}. 
From GALFIT we obtain two parameters, the ellipticity and the effective radius, that tell us quantitatively about the size and morphology of these DLA counterparts.
We list the morphological parameters in table \ref{tab:galfitparameters}.

\begin{table*}
\centering
\caption{Parameters of the GALFIT morphological models on the individual galaxy clumps. For each of these we consider the profile that we fit in the optical filter F606W, except for the case of Q0302$-$223, where we present the values for the F160W filter. The parameters we show are the integrated magnitude AB $mag$, the effective radius $\rm R_e$ in units of pixels, the Sersic index $n$ and the axis ratio $b/a$. }
\label{tab:galfitparameters}
\begin{tabular}{lcccccc}
\hline
Object & filter & AB $mag$ & $R_{\rm e}$ [px] & $R_{\rm e}$ [kpc] & $n$ & $b/a$ \\ 
\hline
Q0302$-$223 (4) & F160W & 23.98 $\pm$ 0.12 & 2.14 $\pm$ 0.25 & 2.3 $\pm$ 0.3 & 0.62 $\pm$ 0.13 & 0.5 $\pm$ 0.05 \\
Q0302$-$223 (3) & F160W & 22.77 $\pm$ 0.06 & 2.29 $\pm$ 0.12 & 2.4 $\pm$ 0.1 & 0.86 $\pm$ 0.08 & 0.82 $\pm$ 0.04 \\
Q0452$-$1640 (1) & F606W & 24.01 $\pm$ 0.12 & 7.60 $\pm$ 0.56 & 2.5 $\pm$ 0.2 & 0.50 $\pm$ 0.11 & 0.36 $\pm$ 0.04 \\
Q0452$-$1640 (2) & F606W & 25.14 $\pm$ 0.17 & 2.40 $\pm$ 0.45 & 0.8 $\pm$ 0.1 & 0.56 $\pm$ 0.26 & 0.83 $\pm$ 0.17 \\
Q1009$-$0026 & F606W & 22.46 $\pm$ 0.05 & 9.95 $\pm$ 0.35 & 3.1 $\pm$ 0.1 & 0.47 $\pm$ 0.04 & 0.88 $\pm$ 0.03 \\
Q2222$-$0946 & F606W & 24.21 $\pm$ 0.16 & 4.62 $\pm$ 0.39 & 1.5 $\pm$ 0.1 & 0.81 $\pm$ 0.21 & 0.38 $\pm$ 0.06 \\
Q2352$-$0028 (1) & F606W & 25.05 $\pm$ 0.18 & 5.79 $\pm$ 0.69 & 1.9 $\pm$ 0.2 & 0.55 $\pm$ 0.19 & 0.49 $\pm$ 0.08 \\
Q2352$-$0028 (2) & F606W & 26.60 $\pm$ 0.35 & 3.32 $\pm$ 0.65 & 1.1 $\pm$ 0.2 & 0.34 $\pm$ 0.27 & 0.43 $\pm$ 0.13 \\
\hline
\end{tabular}
\end{table*}

\begin{table}
	\centering
	\caption{Derived stellar physical properties of the Lyman-$\alpha$ absorbing galaxies from SED fitting. The second column provides the stellar masses. The third and fourth columns provide the baryonic mass fraction, $f_{\rm baryons}$, and the gas fraction, $f_{\rm gas}$, for these objects {as defined in section 5.2}.}
	\label{tab:physprop}
	\begin{tabular}{cccc} 
		\hline
 Quasar Field & log($M_*$)  &$f_{\rm baryons}$ &$f_{\rm gas}$\\ 
&[$\text{M}_{\odot}$] & [percent] & [percent]\\
\hline
Q0302$-$223   & 9.5$\pm$0.2		 		& 22$\pm$8	&28$\pm$30\\
Q0452$-$1640 & 9.1$\pm$0.2  		 	& 7$\pm$2	&56$\pm$20\\
Q1009$-$0026 & 10.7$\pm$0.2   	& 65$\pm$30	&3$\pm$50\\ 
Q2222$-$0946 & 9.7$\pm$0.3   	  	& $>$100	&50$\pm$40\\
Q2352$-$0028 & 9.4$\pm$0.3   & 13$\pm$7 	&20$\pm$60\\
		\hline
	\end{tabular}
\end{table}

\subsection{Spectral Energy Distribution Fit}

\begin{figure*}
	\centering
	\includegraphics[width=0.9\columnwidth]{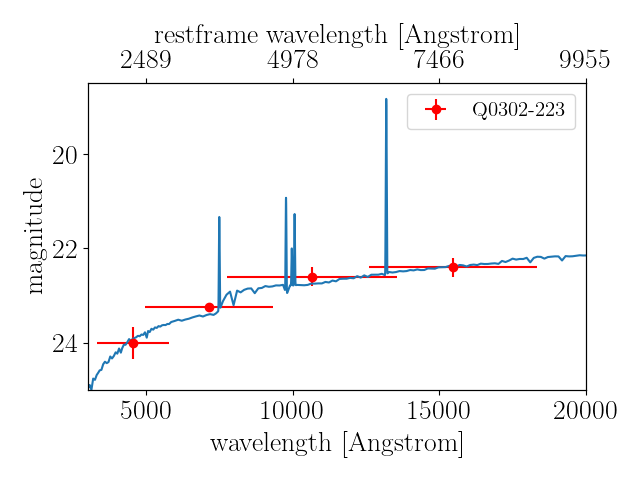}
	\includegraphics[width=0.9\columnwidth]{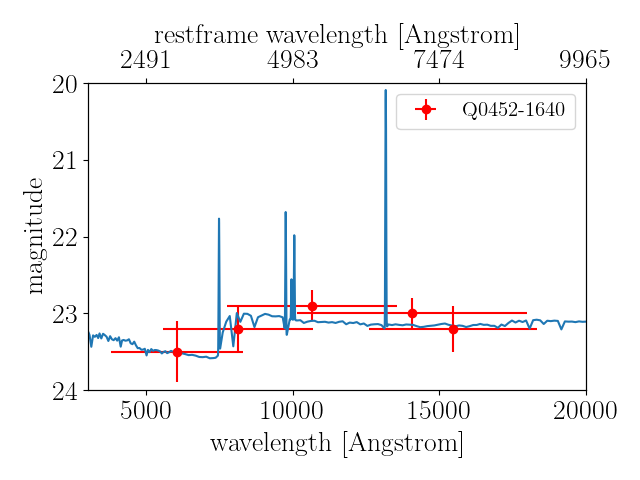}
	\includegraphics[width=0.9\columnwidth]{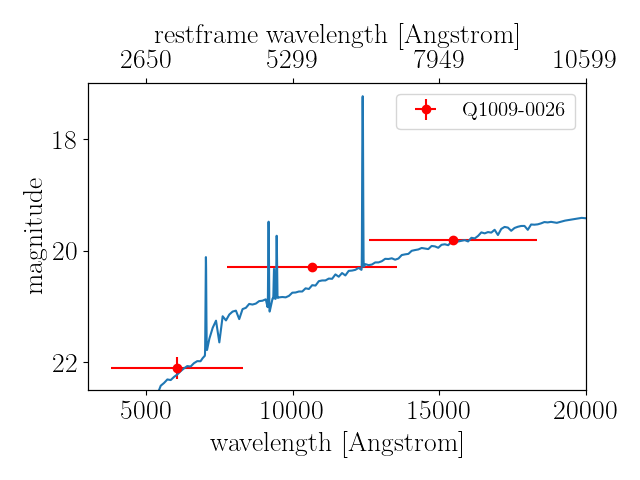}
	\includegraphics[width=0.9\columnwidth]{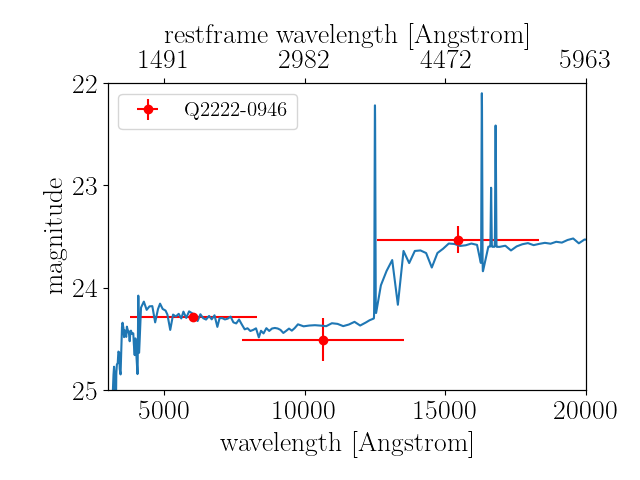}
	\includegraphics[width=0.9\columnwidth]{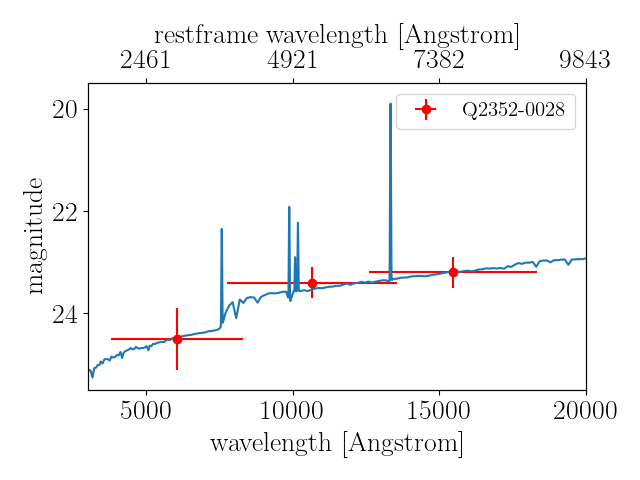}
    \caption{SED fitting to the broad-band magnitudes from HST observations for the Lyman-$\alpha$ absorbing galaxies in the sample. The SED fit {in blue} is based on the LePhare photometric-redshift code \citep{ilbert09}. The spectroscopic redshifts derived from the 
VLT/SINFONI spectra are used as input to the code. 
{The red data points give the measured magnitudes with their errors (in y-direction) and the filter wavelength range (in x-direction).}
The derived stellar masses (see table \ref{tab:physprop}) are reliable, while the age of the stellar population and the SFR are less robustly determined. }
    \label{fig:sed}
\end{figure*}

We use the derived broad-band magnitudes {in table \ref{tab:magnitudes}} which cover the 4000 \AA\ Balmer break of the object to constrain the mass, SFR, and age of the stellar population in the galaxy with a spectral energy distribution (SED) fit to the integrated light of the galaxies. 
We utilised the photometric-redshift code {\it LE PHARE} \citep{arnouts99,ilbert06} available as part of the {\it GAZPAR} suite of software\footnote{https://gazpar.lam.fr/home}. We used the code with a single burst of star formation, templates from \citet{bruzual03} stellar population models, a Calzetti extinction law and a \citet{chabrier03} initial mass function. The spectroscopic redshifts derived from our SINFONI spectra are used as an input to the code, thus allowing us to constrain the stellar mass of the object with relatively high confidence. 
Based on previous experience \citep{peroux11b,ilbert06}, we know that the derived stellar masses are reliable, while the age of the stellar population  and the SFR are less robustly determined. 
{Figure \ref{fig:sed} shows the best fit template for each of the counterpart objects, based on the averaged magnitude from the template spectrum in the range of the respective bandwidths, convolved with the throughput. Note that, while the measured F105W magnitude for Q1009-0026 is noticeably offset from the continuum, this band contains a number of strong emission lines that are taken into account to determine the best fit template.}

We note that three of our galaxies (Q0302$-$223(3+4), Q1009$-$0026 and Q2222$-$0946) already have previously published mass estimates from SED fitting.
\citet{christensen14} used a variety of magnitudes available in literature and the code HYPERZ \citep{bolzonella00} to perform an SED fitting and obtain stellar masses.
For the DLA counterpart towards Q0302$-$223 we find log($M_*/\text{M}_{\odot}$) = 9.5 $\pm$ 0.2, which is consistent with \citet{christensen14} who found 9.65 $\pm$ 0.08.
In Q1009$-$0026 they find log($M_*/\text{M}_{\odot}$) = 11.06 $\pm$ 0.03, which {is somewhat} higher than our result of 10.7$\pm$ 0.2.
Our result for Q2222$-$0946 (log($M_*/\text{M}_{\odot}$) = 9.7$\pm$0.3) is consistent within the errors with their result of 9.62 $\pm$ 0.12.
{\citet{krogager13} have found (log($M_*/\text{M}_{\odot}$) = 9.3$\pm$0.2) for Q2222$-$0946, which is in agreement within the errors.}
In contrast to \citet{christensen14} we have a homogeneous data set with space based observations in three filters that cover the 4000 \AA\ Balmer break, which is critical for the SED fitting.
In addition we apply to the high spatial resolution data a consistent magnitude and aperture correction estimation method.
For these reasons, we focus on the results from our new HST magnitude measurements in the following analysis.

\section{Results and Discussion}

\subsection{The Stellar Mass of Lyman-$\alpha$ Absorbing Galaxies}

\begin{figure}
	\includegraphics[width=1.0\columnwidth]{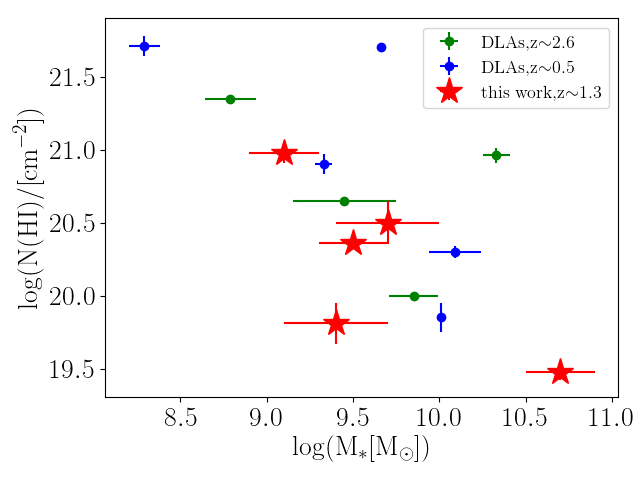}
    \caption{Neutral gas column density, N(\ion{H}{i}) as a function of Lyman-$\alpha$ absorbing galaxies stellar masses. When the absorbing galaxies have been detected and their stellar masses have been measured, DLA and sub-DLAs appear to have stellar masses ranging from 8$<$log($M_*/\text{M}_{\odot}$)$<$11 solar masses. We find an apparent anti-correlation between log($M_*/\text{M}_{\odot}$) and $N$(\ion{H}{i}). We obtain a Spearman rank correlation coefficient of -0.6 for these data points. {The literature points for the high and low redshift samples are taken from \citet{krogager17}.
    The one point from our data, which seems to be an outlier in the lower left, is Q2352$-$0028.}
    }
    \label{fig:massnhi}
\end{figure}

We have measured the stellar mass of the five DLA counterparts in our sample {(see table \ref{tab:physprop})}. 
Together with the objects from literature, about 15 absorbing galaxies have now their stellar mass measured (\citealp{christensen14,krogager17}; Rhodin et al. in prep.). 
Figure~\ref{fig:massnhi} shows the resulting distribution of stellar masses as a function of $N$(\ion{H}{i}) column density.  
Absorbers with log($N$(\ion{H}{i})/cm$^{-2}$) > 20.3 have smaller masses than Milky-Way-like galaxies, {which is expected for most galaxies at z$\sim$1}.
The absorbing galaxies stellar masses appear to be weakly anti-correlated with $N$(\ion{H}{i}) column densities (Spearman rank correlation coefficient: $-$0.6), where higher stellar masses are traced by lower \ion{H}{i}-absorbers. 
This is in line with results from \citet{kulkarni10} who use independent arguments based on metallicity to show that on average sub-DLAs might arise from more massive galaxies than DLAs.
Galaxies with higher metallicity would have undergone a more rapid star formation and gas consumption, leaving them with lower $N$(\ion{H}{i}) in their vicinity.
On the other hand, detection of high mass - high column density systems could suffer from dust extinction within the galaxy itself \citep{vladilo05}, thus affecting the magnitude of the background quasar.
This observational bias could also explain the lack of systems in the right top corner in the figure \ref{fig:massnhi}.

We note one apparent outlier in this relation, which is Q2352$-$0028.
This object lies slightly below the main relation that is formed by the remaining data points.
An explanation is an ongoing interaction in this system (see also discussion about morphology later).
We would expect a more massive counterpart for the measured $N$(\ion{H}{i}).
The tidal interaction in this system could have deposited some \ion{H}{i} gas on top of the quasar sight line and cause the apparent deviation from the $N$(\ion{H}{i})-$M_*$ relation.

The stellar masses we report here span an interval from 9<log($M_*/\text{M}_{\odot}$)<11. 
However, it is important to bear in mind that many DLAs and sub-DLAs counterparts still remain undetected to date, so that the sample might not be representative of the population of absorbers as a whole. 
Recent findings by \citet{krogager17} and the study by \citet{moller13} indicate that DLAs trace a large range of stellar masses with an estimated average stellar mass of log($M_*/\text{M}_{\odot}$) $\sim$ 8.5. 

It is interesting to compare these measurements with the most recent simulations of cold gas. 
Clearly, resolving these mass scales in a cosmological context is a challenge to hydrodynamical simulations, as sub-grid physics are not properly simulated. 
\cite{pontzen08} estimate that DLAs are dominated by galaxies with 7$<$log($M_*/\text{M}_{\odot}$)$<$8. 
\cite{tescari09} and \cite{rahmati14}, find that most \ion{H}{i} absorbers with $N$(\ion{H}{i}) < 10$^{21}$cm$^{-2}$ are associated with very low mass galaxies, log($M_*/\text{M}_{\odot}$) < 8. 
{However}, \cite{rahmati14} found that at higher \ion{H}{i} column densities the contribution of haloes with log($M_*/\text{M}_{\odot}$) > 9 increases rapidly. 
\cite{rahmati14} use the EAGLE simulations to show that at z$\sim$3, sub-DLAs are dominated by systems of 7.0$<$log($M_*/\text{M}_{\odot}$)$<$8.5, while at log $N$(\ion{H}{i})>20.7, objects with 8.5$<$log($M_*/\text{M}_{\odot}$)$<$10 progressively dominate. 
Based on precursors of the Illustris simulations, \cite{bird14} find that the DLA population probes a wide range of halo masses, but that the cross-section is dominated by haloes of mass 10$<$log($M_{\rm halo}/\text{M}_{\odot}$)$<$11 solar masses. 
Overall, the simulations predict lower stellar masses than found in our study.
{This illustrates the many challenges in reproducing this cold phase of the gas in an cosmological context.}

\begin{figure}
	\includegraphics[width=1.0\columnwidth]{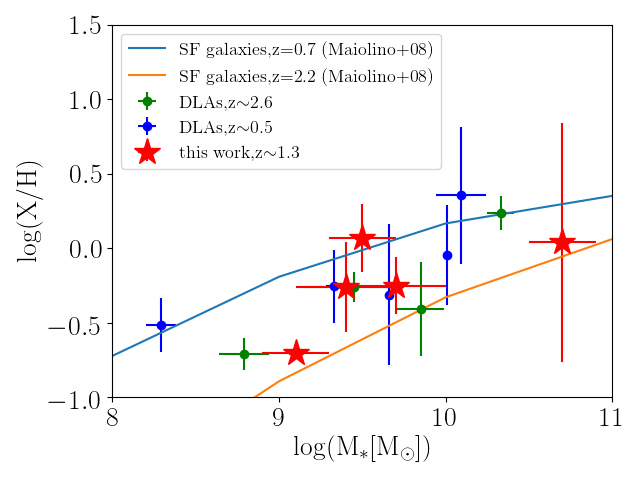}
    \caption{Mass-metallicity relation for absorbing galaxies. The metallicities of the literature samples {(compilation from \citet{christensen14})} in green and blue are neutral gas metallicities measured in {absorption}. 
   {They are corrected for the impact parameter to be comparable to emission metallicities.} Here, we have been able to measure {\it directly} both the stellar masses and emission metallicities of the absorbing galaxies {for our sample in red}, so that assumptions on metallicity gradients are not required.
{We plot the emission metallicities from R23 to be comparable to \citet{maiolino08} except for the cases of Q2352$-$0028, for which the R23 metallicity could not be determined, and Q1009$-$0026, which was not observed with X-Shooter \citep{peroux14}. 
In those two cases we plot the metallicity from N2 (see also table \ref{tab:knowndata}).}    
    }
    \label{fig:mass_metal_corr}
\end{figure}

{On the other hand it might also indicate a detection bias of DLA counterparts.}
Massive galaxies are on average more luminous and easier to detect in the vicinity of a bright background quasar than low mass objects that are fainter on average.
The more massive a galaxy, the larger also its extent and the possible impact parameter of a QSO sight line.
{Also, the more massive a galaxy, the higher its metallicity (see e.g. Figure \ref{fig:mass_metal_corr}).
\citet{krogager17} have shown a trend between impact parameter and metallicity in DLAs which can be translated into a trend between impact parameter and mass.
}
{There is a possibility that some {observations} are biased to detecting systems with larger impact radii because they are easier to separate from the QSO.}
{
However, we detect DLA counterparts down to impact parameters of $\sim$ 6 kpc.}
{Still, the lowest stellar mass we measure in our sample is $\sim 10^9 \text{M}_{\odot}$, which is higher than the bulk of the population predicted by simulations.}

\subsection{The Baryonic and Gas Fractions of Lyman-$\alpha$ Absorbing Galaxies}

Thanks to the wealth of data available for this unique sample, we are able to measure the dynamical, halo, gas and stellar masses of these absorbing galaxies {(see tables \ref{tab:knowndata} and \ref{tab:physprop})}. 
The dynamical masses are calculated from the virial theorem or from the enclosed mass if the system is found to be rotating using a converged $V_{\rm max}$ derived from a 3D fit to the IFU H-$\alpha$ observations (see \citealp{peroux11a} for more details). 
The halo masses are computed from the H-$\alpha$ emission widths and assuming a spherical virialised collapse model. 
The gas masses are indirectly derived from the observed H-$\alpha$ surface brightness using an 'inverse' Schmidt-Kennicutt relation \citep{peroux14}. 

The molecular gas component is often found to be considerably lower than the \ion{H}{i} gas, typically 20 percent of the total gas mass \citep{lelli16}. 
{This gas component has been studied in absorption \citep{noterdaeme15a,noterdaeme15b,balashev17} towards DLAs. 
These studies imply that molecular gas is preferentially found in high \ion{H}{i} column density DLAs (log($\rm N_{\ion{H}{i}}$) > 20.7) and makes up a few percent of the total gas content.
}

Recently, there have {also} been searches for CO emission with ALMA in DLA counterparts as a tracer for molecular gas. 
While there were a number of non detections, \citet{neeleman16} report a molecular gas mass of log($M_{\rm mol}$/M$_{\odot}$) = 9.6 for one of the objects in the \citet{christensen14} sample at low redshift (PKS 0439$-$433, $z$=0.101, log(N(\ion{H}{i})/[cm$^{-2}$])=19.63, log($M_{*}$/M$_{\odot}$)=10.01). 
Similarly \citet{moller18} report a molecular gas mass in a redshift 0.7 DLA (log($M_{*}$/M$_{\odot}$)=10.80) of log(($M_{\rm mol}$/M$_{\odot}$) = 10.37.
\citet{klitsch18} report several {galaxies} associated with a $z$=0.633 Lyman Limit System (LLS). 
One of these is detected in multiple CO lines leading to log($M_{\rm mol}$/M$_{\odot}$) = 10.1.
The non-detections, given their stellar masses (10.2 $<$ log($M_{*}$/M$_{\odot}$) $<$ 11.2), indicate that the molecular gas mass makes up only a small fraction of the total baryonic mass in these galaxies.
For one of the absorbing galaxies studied here (Q2222-0946) we have a direct 5-sigma upper limit on the molecular mass from CO(3-2) ALMA non-detection of log($M_{\rm mol}$/$M_{\odot}$) < 9.89, which corresponds to < 50 percent of the gas being molecular.

The available mass estimates allow us to put constraints on the baryonic mass fraction in the DLA galaxies:

\begin{equation}
f_{\rm baryons}=(M_{\rm gas}+M_{*})/M_{\rm dyn}
\end{equation}
 
The baryonic fractions are tabulated in table~\ref{tab:physprop} and are found to vary vastly from one object to another. 
\citet{lelli15} find high values (60 $-$ 100 percent) for local tidal dwarf galaxies (log($M_*$/M$_{\odot}$) $\sim$ 8). 
The two objects for which we measure baryon fractions in this range, have relatively large stellar masses of log($M_*$/M$_{\odot}$) = 9.7 and log($M_*$/M$_{\odot}$) = 11.6.
It is unexpected that the two most massive systems in our sample fall into the baryonic fraction range of tidal dwarf galaxies, while the other systems that are closer to tidal dwarfs in terms of their masses and their morphology (see also section 5.4) appear to have much lower baryon fractions.

Similarly to the baryonic fractions, we can estimate the gas fraction in these absorbing galaxies:

\begin{equation}
f_{\rm gas}=M_{\rm gas}/(M_{\rm gas}+M_{*})
\end{equation}

We derive gas fractions ranging from a few percent (in the case of the absorbing-galaxy in the field of Q1009$-$0026) to 56 percent (for Q0452$-$1640). Such gas fractions are in the low range of the typical values derived in $z\sim$2-3 galaxies by others (e.g. \citealt{erb06gas,law09}). 
\cite{bahe16} make predictions of the neutral gas fraction as a function of stellar mass at $z$=0 (10 $< M_*$/M$_{\odot}$ $<$ 11.5). 
Our most massive galaxy (Q1009$-$0026) is the only one from our sample that falls into their mass range.
It agrees with their predictions, although it falls into the lower range of the predicted neutral gas fraction.

\begin{figure}
	\includegraphics[width=1.0\columnwidth]{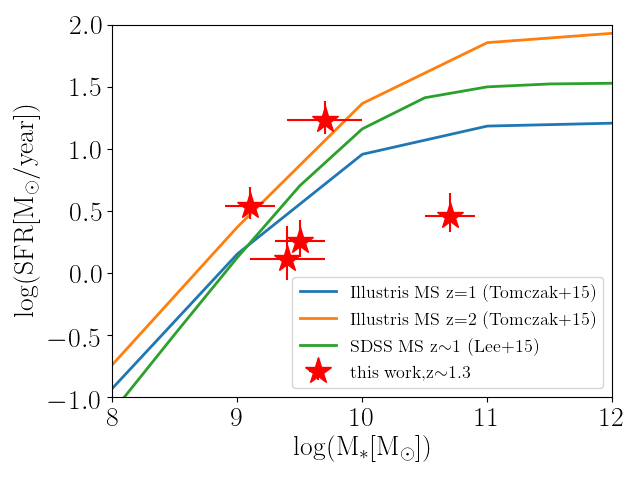}
    \caption{H-$\alpha$ star formation rates of our targeted absorbing galaxies as a function of their stellar masses. We compare our results with the predictions for the main sequence (MS) of star forming galaxies from the Illustris simulation \citep{tomczak16} as well as with the main sequence found from SDSS \citep{lee15}. Except for the most massive one, our data points are broadly speaking in line with these relations, implying that DLA counterparts are normal star forming galaxies and do not form a special group of galaxies.}
    \label{fig:mainsequence}
\end{figure}

\subsection{Lyman-$\alpha$ Absorbing Galaxies Scaling Relations}

In this section we will investigate some stellar-mass scaling relations.

{\em The Star-Forming Sequence:}
Figure~\ref{fig:mainsequence} shows the H-$\alpha$ star formation rates of our targeted absorbing galaxies as a function of their stellar masses. 
For comparison, we plot the SDSS Main Sequence at $z$=1 from \cite{lee15} as well as results from the Illustris simulations at $z$=1 and $z$=2 \citep{tomczak16}. 
Our data points are broadly speaking in line with these relations, except for the most massive object in our sample.
They do not systematically deviate from the expected relation for star forming galaxies.
This implies that DLA counterparts do not form a special group of galaxies, but follow the relations of normal star forming galaxies.
{The most massive galaxy seems to fall below the main sequence although the star formation rates are not dust corrected.}
\\

{\em The Mass-Metallicity Relation:}
\cite{peroux03} reported a correlation between the absorption metallicity of the absorber and the absorption profile velocity spread along the quasar line of sight, $\Delta v$. \cite{ledoux06} later argued that $\Delta v$ could be a proxy for the mass of the systems and advocate that the relation is an analogue of the mass-metallicity relation (MZR) of normal galaxies. 
\cite{moller13} further used this relation to predict the stellar mass of absorbing galaxies based on their metallicities, while \cite{christensen14} assumed a perfect MZR to derive the metallicity gradients of these systems. 
Figure~\ref{fig:mass_metal_corr} shows where our data points lie on this relation.
We plot the emission metallicities for the objects in our sample and the impact-parameter-corrected absorption metallicities for the literature sample \citep{christensen14}. 
{This impact-parameter-correction assumes a constant metallicity gradient from the center of the galaxy outwards such that absorption metallicities can be related to a central emission metallicity. 
This enables one to compare absorption selected samples with emission selected samples.}
For comparison, we have plotted as solid lines fits for star-forming galaxies at $z$=0.7 and 2.2 \citep{maiolino08}.
We find our data points in agreement with the previously determined MZR for DLAs.

Here, we have been able to measure {\it directly} both the stellar masses and emission metallicities of the absorbing galaxies, so that assumptions on the relation between $\Delta v$ and mass on one hand and metallicity gradients are not required.

\begin{figure}
\includegraphics[width=1.0\columnwidth]{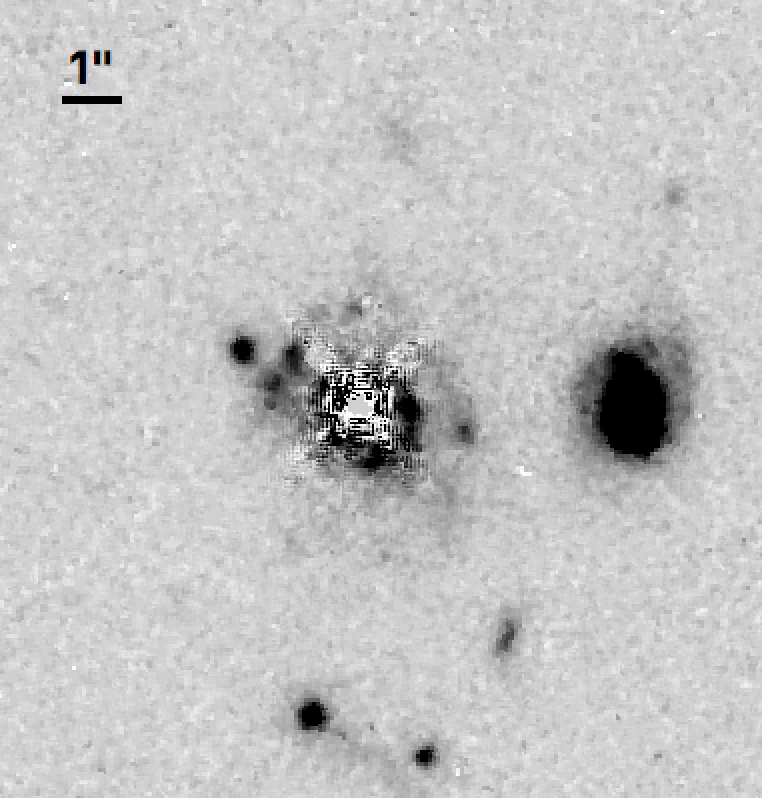}
\caption{Field of Q1009$-$0026 (Proposal ID: 13733, PI: P{\'e}roux): The prominent spiral galaxy on the {west side of the quasar} is the object that we identify as the DLA counterpart.
{The clumpy spiral structure which is very obvious in the optical band (see figures \ref{fig:sample} and \ref{fig:morphologyandgalfit}) is less prominent in this near-IR image.} 
Besides this counterpart galaxy we find a lot of structure around the quasar line of sight after PSF subtraction. See text for further discussion. North is up and East is left. The filter shown here is F105W.}
\label{fig:1009field}
\end{figure}
\subsection{The Morphological Complexity of Lyman-$\alpha$ Absorbing Galaxies}

In this section, we provide further details on the morphology of the Lyman-$\alpha$ absorbing galaxies that have become obvious with the high spatial resolution provided by the HST imaging. 
Figure~\ref{fig:morphologyandgalfit} provides for each field, the observed, modeled and residual from our detailed GALFIT modeling in the optical filter F606W or for F160W in the case of Q0302$-$223. 
The discussion is mainly based on these, as we have the highest sampling with a pixel scale of 0.04 arcsec/px in the optical filter F606W and the best detection of Q0302$-$223 in F160W.
{The residuals in Figure~\ref{fig:morphologyandgalfit} show the deviation of the observed galaxies/clumps from the ideal Sersic profile and provide evidence for sub-structure within the observed object.
The appearance of clumps within galaxies is generally related to either some external disturbance such as interactions or due to instabilities within a disk and is dependent of the spatial resolution of the observations.
The extreme examples in terms of resolution are Q1009$-$0026 at the lowest redshift of $\sim$0.9, which shows a lot of clumpy residuals and Q2222$-$0946 at the highest redshift of $\sim$2.4, which presents the smoothest profile.
In the following, we discuss the morphology of the galaxies in each individual field:
 }

{\em The Q0302$-$223 Field:}
In this field we find two close galaxies at the position of the H-$\alpha$ emission (3+4, {see figure \ref{fig:label})}. 
These two galaxies had already been found previously as well as two other clumps (1+2) that are even closer to the quasar sight line.
Since the later two (1+2) did not show any H-$\alpha$ emission at the redshift of the absorber in the SINFONI IFU data, we assume that the galaxies responsible for the Lyman-$\alpha$ absorption are objects 3 and 4.
Due to their proximity we may assume that they are in an interaction with each other although we do not see any features that hint at some interaction like tidal streams.
We can not see anything fainter than a magnitude of $\sim$ 26, so if there are some very faint tidal streams, we might miss them in our observations.
It should be noted at this point that we discuss the morphology for this field in a different filter (F160W) than the other objects in our sample (F606W) and might therefore probe slightly different stellar populations. 
For the scope of this qualitative discussion we ignore the expected minor changes in morphology due to different stellar populations.

{\em The Q0452$-$1640 Field:} 
{This galaxy has the clear profile of a single rotating disc in the SINFONI data.}
The HST images of this target show an elongated and a compact object. 
This might hint at two merging objects but since we see a clear disc in the infrared and in H-$\alpha$ there is a strong possibility for this object to be an edge-on disc {(b/a=0.36)} galaxy with two separate clumps of star formation or a patch of dust that absorbs the emission in the optical.
\citet{elmegreen12} have investigated so-called tadpole galaxies that typically have a strongly off-centered region of star formation and an elongated intensity profile. 
The DLA counterpart that we find here might be an example for such a tadpole galaxy.

{\em The Q1009$-$0026 Field:} The HST images show the presence of clumps and spiral arms. After quasar PSF subtraction, we also note a number of smaller objects at small impact parameters to the quasar (see figure \ref{fig:1009field}). We measured the magnitudes for these but due to their proximity to the quasar, the errors are large and we cannot get a constraint of the redshift of these clumps from SED fitting.
\citet{meiring11}, {who had evidence for two absorption systems towards this quasar,} were looking for a second sub-DLA counterpart in this field but could only identify a faint cloud south of the QSO.
In our high-resolution imaging data we can see that there are in fact several objects both south as well as in the vicinity of the QSO.
Lacking further information on the redshift of these objects we cannot identify the second sub-DLA counterpart at $z$=0.8426 . 

{\em The Q2222$-$0946 Field:}
We find this object to be the most compact and isolated within our sample.
It does not show any companions or a specific structure.
It might therefore be a fully evolved field galaxy, {although it lies at high redshift}.
Since it is also the one with the highest redshift ($z$=2.35) it is not as highly resolved as the other ones at lower redshift and we might not be able to resolve individual components.
{It is also possible that we probe sightly different stellar populations in this galaxy than in the others due to its higher redshift and therefore more redshifted spectra.}

{\em The Q2352$-$0028 Field:} The HST images show a disturbed morphology pointing towards the quasar. 
We discover clear tidal streams connecting the two clumps and also stretching towards the QSO sight line.
This larger sky coverage of {tidally stripped gas favours detecting interacting systems in absorption}.
Some tidal streams stretch out from the host galaxies and cause the strong absorption feature in the QSO spectrum.
Without these tidal features, the CGM around galaxies {could} be less extended.
{As \citet{hani18} have shown, galaxy mergers can enhance the extent of the CGM.}

When comparing the clumpiness to the baryon fractions presented in table \ref{tab:physprop}, we find an apparent dependence of the baryon fraction on the structure of the galaxy.
For those with only a single galaxy (Q1009$-$0026 and Q2222$-$0946) we find extremely large baryon fractions (65-100 percent).
The DLA counterparts that show a clumpy structure that might hint at an ongoing merging are found to have low baryon fractions (7-22 percent).
These might in fact be a further indication of ongoing merging as the tidal interaction between the clumps might cause the baryonic matter to be scattered throughout the halo out to large radii, leading to lower baryon fractions in the central part of the halo.
The fully evolved galaxies on the other hand accumulate all of their stars and gas in  the central region of the dark matter halo and have therefore a core that is dominated by baryonic matter.

The absorption reaches up to 40 kpc outside the galaxy center. 
This means that some processes must drive this gas far out of the galaxy and keep it there.
In the framework where many of these DLA galaxies actually consist of interacting clumps, some of them are maybe filaments that were created by {tidal interactions between clumps}.
This means that the gas we see in absorption is rather associated to a group of smaller clumps than to a single galaxy, {which is in line with findings from simulations \citep{rahmati14}.}
The observations from \citet{fumagalli14b} and \citet{cantalupo14} also suggest that the CGM {around quasars} is composed of streams and filaments.

{
Already the first identified DLA galaxy \citep{moller93} was found in a group comprising at least 3 galaxies. 
Later, spectroscopy \citep{warren96} revealed the tight kinematics of this group with an estimated merging timescale of $\rm < 10^{9}$ years.  
Later work has confirmed that absorption selected galaxies often are found in similar tight group environments, or even in the process of active merging. 
In the sample of \citet{christensen14} at least five of the 12 DLA galaxies are either in groups or parts of actively merging systems.  
More recent work \citep{peroux16,bielby17,fumagalli17,rahmani18,klitsch18} has confirmed the high fraction of group environments encountered in the search for  galaxies with strong absorption features. 
While our ground based SINFONI data originally suggested that these five DLA counterparts are isolated galaxies, we found that three out of five of our objects have substructure in the HST data, which we call clumps.
This further suggests that the strong absorption that we see towards quasars are preferentially caused by some group-like structure that contains interacting objects and intergalactic material.
}

In conclusion, the enhanced spatial resolution provided by HST imaging reveals the complexity of the morphology of these Lyman-$\alpha$ absorbing galaxies. 
We find that the DLAs and sub-DLAs are associated with isolated, {clumpless} galaxies only in a few cases. 
More often, the absorbing galaxies are within group structures or show indications of interacting clumps (Q0302$-$223, Q0452$-$1640, Q2352$-$0028). 
These results have important consequences for the interpretation of the gas content and metallicity gradient of the circum-galactic medium of galaxies selected by their absorption signature.

\section{Conclusion}

We have analysed the stellar continuum of five DLA counterpart galaxies using HST broad-band imaging in the optical and near infrared (F606W, F105W, F160W).
We measured the stellar masses of these galaxies from an SED fit to the magnitudes we obtained from the HST data.
From our results presented in section 5, we draw the following conclusions:

\begin{itemize}

\item We find an anti-correlation between $N$(\ion{H}{i}) and the stellar masses. 
 \citet{kulkarni10} proposed that sub-DLAs trace more massive galaxies than DLAs.
The former could have undergone more rapid star formation and gas consumption leading to lower $N$(\ion{H}{i}) in their surrounding medium.
This anti-correlation could also be due to an observational bias as high-mass-high-$N$(\ion{H}{i}) systems could be more difficult to detect because of dust obscuration.

\item Using these stellar masses and the estimates available for the gas mass and the dynamical mass (see table \ref{tab:knowndata}), we calculate the gas fraction and the baryonic fraction within the observed galaxies.
We find a large spread {(7-100\% for baryon fractions, 3-56\% for gas fractions)} in these values among our sample.

\item Combining these observations with emission spectra, we investigated the scaling relations of DLAs.
{The majority} of our Lyman$-\alpha$ counterparts follow the main sequence of star forming galaxies, given their stellar masses.
They also follow the Mass-Metallicity Relation for DLAs.

\item Making use of the high spatial resolution offered by the HST imaging, we investigate the morphology of these DLA counterparts.
{Most are not fully} evolved disk galaxies, but rather composed of individual star forming clumps that are in close interaction.
Especially in the optical broadband filter F606W which has the highest spatial resolution among our data sets, individual clumpy structures instead of a single disk galaxy are found.
{Three out of five galaxies show clumpy structure, while the remaining two could be disks.}
When fitting Sersic models to the observed galaxies we find residuals that show filamentary structure, indicating gas flows due to interaction of different clumpy components.
\end{itemize}

Taking these points together, the circum-galactic medium of DLA galaxies appears complex.
A larger sample of DLA counterparts combining absorption observations with CGM in emission are required to gain a complete understanding of the gas flowing processes in and around  galaxies.

{In addition to the optical searches such as the one described here, there are on-going campaigns detecting with ALMA molecular emission lines from absorption selected galaxies \citep{neeleman16,neeleman18,klitsch18,moller18,kanekar18}.
Combining optical and sub-mm wavelengths observations of absorber host galaxies will bring new insights into the role of molecular gas in the CGM of galaxies.}

\section{Acknowledgements}
RA acknowledges the ESO and CNES studentships.
CP thanks the ESO science visitor programme and the DFG cluster of excellence `Origin and Structure of the Universe'.
VPK acknowledges partial support from (NASA) grant NNX14AG74G and NASA/Space Telescope Science Institute support for Hubble Space Telescope program GO-13733 and 13801. 
The data presented in this paper were obtained from the Mikulski Archive for Space Telescopes (MAST). STScI is operated by the Association of Universities for Research in Astronomy, Inc., under NASA contract NAS5-26555. Support for MAST for non-HST data is provided by the NASA Office of Space Science via grant NNX09AF08G and by other grants and contracts.
This paper makes use of the following ALMA data: ADS/JAO.ALMA\#2015.1.01131.S. ALMA is a partnership of ESO (representing its member states), NSF (USA) and NINS (Japan), together with NRC (Canada), MOST and ASIAA (Taiwan), and KASI (Republic of Korea), in cooperation with the Republic of Chile. The Joint ALMA Observatory is operated by ESO, AUI/NRAO and NAOJ.
\bibliographystyle{mnras}
\bibliography{hstpaper} 
\end{document}